\documentclass[aps,floatfix,prd,showpacs]{revtex4}
\usepackage{graphicx}
\usepackage{dcolumn}
\usepackage{bm}

\begin{document}

\title{Magnetic properties of photosynthetic materials - a nano scale study}
\author{Abhishek Bhattacharya$^{1}$, Sufi O Raja$^{1}$, Md. A Ahmed$^{2}$, Sudip Bandyopadhyay$^{2}$, Anjan Kr. Dasgupta$^{1}$}
\affiliation{$^{1}$Department of Biochemistry, University of Calcutta, 35, Ballygunge circular road, Kolkata 700019, India
$^{2}$Department of Physics, University of Calcutta, Rajabajar Science College, 92 A.P.C. road, Kolkata-700009, India.}

\date{\today}

\begin{abstract}
Photosynthetic materials form the basis of quantum biology. An important attribute of quantum biology is correlation and coherence of spin states. Such correlated spin states are targets of static magnetic field. In this paper, we report magnetic properties and spectroscopically realizable static magnetic field effect in photosynthetic materials. Two classes of nano-scale assembly of chlorophyll (NC) are used for such a study. Magnetic measurements are made using a superconducting quantum interference device (SQUID). Both ferromagnetic and superparamagnetic states are observed in NC along with a blocking temperature around 250 K. Low temperature quantum (liquid nitrogen) spectroscopy is employed to see how optical transitions are affected in presence of static magnetic field. Plausible practical application aspects of magnetic properties of this optically active material are discussed in the text.
\end{abstract}
\pacs{PACS numbers go here. These are classification codes for your  research. See {\tt http://publish.aps.org/PACS/} for more info.}
\maketitle
\section{Keywords}
Photosynthetic Material, Nanoscale, Magnetic Property, Blocking Temperature, Superparamagnetism, Ferromagnetism, Spin, Nano chlorophyll (NC), Magnetic Memory, FACS, NIRF, Nano Materials.
\section{Introduction}

Photosynthetic pigment, Chlorophyll$\textit{a}$ (Chl$\textit{a}$) acts as a primary photon receptor molecule for light harvesting and photo chemistry. Additionally, multiple  other components such as chlorophyll $\textit{b}$ (other forms are also reported) (1), accessory pigments like beta-carotene, pheophytins, xanthophylls, secondary derivative molecules (protoporphyrins, chlorins, pyrochlorophylls, etc.), proteins (D1, D2, psb, RubisCO etc.), protein bound pigment complexes in membranes (LHC) satisfy the photosynthetic light reactions to collect (like an antenna), transfer (near unity rate) and preserve the photon energy. Chl $\textit{a}$ is an aromatic and macro-cyclic natural dye. Central magnesium (Mg$^{2+}$) atom is linked to nitrogen atoms of the poly pyrrole ring to serve as a site for the excitation energy distribution and control for the excitation transfer reactions (2-3). These hydrophobic molecules tend to aggregate among themselves in contact to water (4-7) and interact among themselves by weak $\pi$-stacking interactions in polar environments (8-13).

Nano scale stabilization and interaction study of similar chemical dyes exemplify the significance and importance of such a photo receptor (14-20). In the present context, the impact of nano scaling and colloidal stabilizations on the properties of the fluorophore molecule and its interaction study had been reported. Interestingly, it may be noted that novel properties may indeed be emergent at the  nano scale (21) such as a crystalline CoSi nano wire exhibit emergent ferromagnetism at the nano scale contrary to its diamagnetic bulk phase (22). 

Magnetic transitions and memory are generally attributed to and are intrinsic to the magnetic materials such as metals (Fe, Co, Ni), magnetic nanoparticles, magnetically susceptible metal nano structures (nano wires (23), magnetic core-shell nanoparticles etc.). However, some reports infer to a pi-stacking dependent magnetic transitions (24-26). Additionally, establishment of the fact that pi-pi interaction is strong enough to beat the cationic repulsion in metalloporphyrins also support the postulate (27) and may mediate assembly stabilizations (28-29). While size dependent alterations in the magnetic transitions of any material is reported to follow the nano scaling laws (30-31), it may be interesting to study how the control over size of such a macro-cyclic fluorophore stabilized as a nano-bio assembly impact to its intrinsic properties and spin chemistry (32).

The primary question addressed in this paper is that, can we replicate magnetic memory effects in cell free systems such as, nano-assembly of chl $\textit{a}$. The question concerns about the mechanism for the magnetic memory component of such an assembly, if any. Instead of looking at a continuous and prior exposure to field, we changed our approach in which we make zero fields and with field magnetic measurements, which would directly predict the magnetic nature of such an assembly. Self-assembly in turn could create grounds for the stabilization of long range magnetic domains, prospering the possibility for the emergence of memory. Affirmative answer to this question would in turn opens up possible implications of this versatile sensory tool to address novel optical and magnetic problems and materials characterization studies. Potentiality for the development of novel opto-magnetic technologies and coherent methods had also been discussed. Lastly, applicability of the NC as a flow cytometry compatible nano-bio and imaging probe had been examined. 

\section{Materials and methods}
\textbf{Isolation and purification of Chl}
Fresh mesophyll tissue from the green leaves of spinach (Spenacea oleracea) was collected for pigment extraction. The tissue are then frozen in liquid nitrogen (-196$^{o}$C) for freeze drying and grinding of the leaves. The pigments were extracted in 2:1:1, methanol: petroleum ether: diethyl ether solvent system. All the procedure was conducted at dark and or dim light conditions. Upper ether layer containing most of the pigments (Chlorophylls and accessory pigment molecules such as pheophytin, carotenoids, xanthophylls, chlorins etc.) was collected carefully. The lower methanol layer containing pigment-protein complexes, membrane remnants, accessory pigments and debris are discarded. Purification of chl $\textit{a}$ from a mixture of pigment analytes was performed using simple chromatographic techniques. Chromatographic separations of the crude pigment mixture was conducted by passing it through a vertical silica gel column (silica gel, SRL India, 100-200 mesh) and/or by thin-layer chromatography (TLC silica gel 60254 or DC kieselgel 60F254 Merck, USA). A liquid/gas phase mixture of mobile phase consisting of n-hexane, n-hexane: acetone (1:1 to 1:0.1. respectively), acetone and acetone: methanol (v/v) had been used for column chromatography (flow rate = 1 ml/min) and a volatile mixture of 60\% petroleum ether, 16\% cyclo-hexane/ n-hexane, 10\% ethyl acetate, 10\% acetone and 4\% methanol (v/v) had been utilized as a mobile phase in case of planar chromatographic separations (33). The $R_{f}$ values for each of the pigments had been determined and the value for free chl was found to be 0.4. Major experimental data were validated by comparing it to a commercial spinach chl $\textit{a}$ from Sigma-Aldrich, USA. Chl $\textit{a}$ is extremely sensitive and prone to degradations to its secondary derivative molecules, photo bleaching, oxidation and thermal damage.

\textbf{Nano chlorophyll (NC) stabilizations}
Chl $\textit{a}$ at a solvent perturbed bulk or free state are phase transferred to absolute polar solvent water. The tetra-pyrrole moiety of the chl $\textit{a}$ tends to self-aggregate by securing the hydrophobic core by $\pi-\pi$ stacking self-interactions. Size controlled nanoparticle stabilization follows as a result of such solvent phase perturbation under ultra-sonication. Citrate was used as a reducing, stabilizing and capping agent for size controlled nanoparticle synthesis. Precipitation of the macro-cyclic aggregates occurs frequently in water without any stabilizing or capping agent. Different concentration of tri-sodium citrate was used to stabilize chl $\textit{a}$ in water based buffers and for the synthesis of desired size of chl $\textit{a}$ nanoparticle (NC). 1 $\mu$g/ml purified chl $\textit{a}$ dissolved in organic solvent was added drop wise to the tri sodium citrate solution placed in a hot water bath (set at boiling point of the organic solvent) under continuous sonication with a probe sonicator. A flow rate of 0.1.5 ml/min was maintained. The sonication was implied with an amplitude frequency set at 80 with 0.5 cycles of interval for 10 to 30 minutes and the resulting colloidal solution containing stable nanoparticle NC were stored in freezer for further characterization studies and applications (34). The hydrodynamic diameter (size) and the relative fluorescence emission of the colloidal NC were found to rely directly on the size of the particles, concentration of the stabilizer and pH of the medium. NC was then diluted in biological buffers such as phosphate buffer (pH 7.2.), DMSO for further experimental measurements. Notably, chl $\textit{a}$ is prone to degradations by thermal damage, oxidation etc. and exert frequent photo-bleaching whereas the colloidal NC is photo stable at a liquid state or at an immobilized film state. 

\textbf{Colloidal stability measurements and zeta potential}
The colloidal stability of the synthesized nanoparticle (NC) was measured in a Beckman Coulter zeta instrument with metal electrode cuvette. The instrument measures the surface potential using an electrophoretic mobility shift assay.

\textbf{UV-Vis spectroscopy}
Absorbance measurements were conducted in a Thermo-Vision Evolution 300 spectrophotometer. A spectral range of 350 nm to 800 nm was scanned for chl a. A xenon lamp was illuminated for absorbance measurements. Nitric oxide sensing and dose dependent experiments were conducted using 96-well plates and plate readers which acquire data at a fixed $\lambda$ corresponding to the O.D$_{max}$ of NC at 25$^{o}$C. 

\textbf{Fluorescence measurements}
Fluorescence measurements were performed in a PTI fluorescence spectrophotometer (Quantamaster$^{TM}$40, USA). Excitation wavelength was set at 430nm with excitation and emission monochromators and emission was collected to a perpendicular direction between 650nm to 750nm for chlorophyll $\textit{a}$. The bandwidth was set at 5nm. Time kinetics data was collected at fixed excitation and emission wavelengths set at 430nm and 665nm respectively. The temperature dependent experiments were performed with a temperature controlled peltier system. Synchronous fluorescence measurements were conducted in a 350nm to 750nm wavelength region with simultaneous excitation and emission scanning. A range of different solvent interacting environments (polarity shifts) and their effect on the fluorescence was examined for purified chl $\textit{a}$. Polarization study was conducted utilizing the corresponding excitation and emission polarizer placed between the light paths.

\textbf{Cryogenic fluorescence measurements}
Low temperature fluorescence emission spectroscopy was conducted in a Hitachi F-7000 spectrophotometer, Japan. The excitation monochromator was set to 480nm (slit width = 5 nm). Emission was scanned between 650 and 800 nm (slit width = 10 nm). Free chl dissolved in solvent and NC at phosphate buffer (pH 7.2) was diluted in presence of cryoprotectant glycerol ($>$60\%) before cryogenic measurements. 77K fluorescence spectra were baseline corrected and the fluorescence ratios for $P_{723}/P_{700}$ was calculated and plotted against wavelength. Fluorescence time kinetics data was acquired at a fixed excitation (480nm) and emission (700nm or 723nm) wavelength against time. Again, a Mg-porphyrin specific excitation of 590nm result in multiple emission bands adjacent to the near-infrared region (NIRF).

\textbf{Dynamic light scattering (DLS/PCS)}
DLS was used as the primary characterization tool for the determination of size and shape of the synthesized nanoparticle NC. The instrument measures the hydrodynamic properties of the nano scale particles depending on their scattering profile and diffusion behavior. Measurements was performed in a Malvern Nano ZS80 (UK) dynamic light scattering set up equipped with a 532nm excitation laser source. All the measurements were conducted at 4$^{o}$C and 25$^{o}$C as mentioned in the text accompanied by a peltier system. The auto-correlation profile (g(2)-1) against the correlation delay times ($\mu$S) infer the liquid state diffusion pattern and the number of scatterers present at a time point. 

\textbf{Atomic force microscopy (AFM)}
Surface topographical imaging of the nanoparticles (NC) was conducted in Nanoscope IVa (Vecco/Digital Instruments Innova, Santa Barbara, CA, USA). Soft silicon probes (RTESPA) was used with a tip radius of 8 nm, mounted on a single-beam cantilever. Cantilever (115-135$\mu$m) deflections were recorded with a cantilever frequency (f0) of 240-308 KHz, horizontal scan rate of 1.2 Hz and 512 samples per line. Spring constant of the cantilevers was set at 20-80 N/m. Scanning was conducted at 25$^{o}$C in air. Data was analyzed by Nanoscope software (Version 5.1.2r3). Phosphate buffer (pH 7.2) soluble NC was then immobilized on to a hydrophilic glass chip. The chips were pre-treated with piranha solution in heat bath for surface functionalization of the glass surface with hydroxyl group (-OH). 

\textbf{Scanning electron microscopy (SEM)}
Scanning electron microscopy was conducted in an Evo18 special edition Carl Zeiss system with EHT of 15kV. The samples were pre-fixed and dried in acetone washed glass chips. The samples were gold sputtered (4nm x 2) before measurements to induce surface conductivity to the samples. 

\textbf{Transmission electron microscopy (TEM)}
Transmission electron microscopy was conducted in a JEM 2100 HR-TEM system with EHT of 200kV. The samples were pre-fixed and dried in a copper grid and were negatively stained with uranyl acetate. The sample was cooled to liquid nitrogen temperatures during the measurements to protect it from burning by harsh electron beam energy.

\textbf{Magnetic measurements (SQUID)}
Magnetic characterizations, field and temperature dependent magnetization measurements was performed in a magnetic properties measurement system (MPMS), superconducting quantum interference device, vibrating sample magnetometer (SQUID VSM, Make: Quantum Design). Chl a was air dried in a heat chamber and NC was lyophilized to dryness before magnetic measurements. The dry weight of the samples was noted for further calculations.

\textbf{Fluorescence spectroscopy in presence and absence of static magnetic field}
A moderate to low strength static magnetic field (SMF) of 100 to 500mT was exposed to the samples. Notably, the static magnetic exposure was imparted before experimental measurements and not during the measurements. 10 minutes of magnetic (SMF) incubation at 4$^{o}$C was imparted to the samples before steady state and time kinetic fluorescence measurements at room temperature as well as at cryogenic temperatures. A 0.01 to 0.05 $\mu$g/ml of sample was diluted in relevant dissolving medium (solvent or buffer) and used for further study as mentioned in the text. Higher concentrations of the fluorophore often exhibit self-quenching effects and hence lower concentrations are selected for the experiments.

\textbf{Relevant objects for characterizations study}
\paragraph{GNP synthesis}
GNP synthesis was carried out using an accepted method (35), with minor modifications (36). An aqueous solution of HAuCl$_{4}$ (20 mM, 25ml) was brought nearly to boiling condition and stirred continuously with a stirrer. Freshly prepared tri-sodium citrate solution (38.8.mM) was added quickly at a time. The citrate concentration is related to the particle size, resulting in a change in color from pale yellow to deep red. The temperature was brought down to normal and the colloidal solution was stirred for an additional few minutes with excess citrate for volume make up. Typical plasmonic resonance for GNP was found at 530nm, which according to standard literature corresponds to a size close to 40nm. The final atomic concentration of gold was calculated to be 200$\mu$M. 

\paragraph{SNP synthesis}
Silver nanoparticle was synthesized by reducing silver nitrate 20 mM (AgNO$_{3}$) solution in presence of sodium borohydrate (100mM). Citrate acts as a stabilizer in the preparation. The stirring was performed at normal temperatures leading to a change in the color of the solution from transparent to yellow rendering SNP formation. After 15 minutes of stirring at room temperature the prepared SNP was incubated in ice for half an hour before use. The plasmon resonance was found in the range 420nm, corresponding to a size 40nm. The final atomic concentration of silver was calculated to be 200 $\mu$M.

\paragraph{Carbon nano materials}
MWNT and amorphous graphene was prepared as described previously (37). A minuscule amount of dry and powered carbon nano material or a 1$\mu$g/ml amount of carbon nano tube and graphene was examined.

\paragraph{Nitric oxide}
Pure nitric oxide donor DETA-NONOate (Cayman Chemicals, USA) was prepared as stock solution of 100 mM in 0.1.(N) NaOH or phosphate buffer saline (PBS; pH 7.4). GSNO was prepared freshly by mixing 1M aqueous solution of sodium nitrite (NaNO$_{2}$, Merck, USA) with 1M solution of reduced glutathione (GSH; Sigma Chemical Co., USA) in 1(N) HCl (HCl, Merck, USA) in 1:1 (v/v) ratio. GSNO formed was protected from light and placed on ice immediately. Further working dilutions (nM, $\mu$M and mM ranges) was prepared freshly before experimental measurements. Sample was incubated with desired amount of sources for 15 to 30 minutes at 37$^{o}$C before measurements.

All the other reagents and chemicals used are of analytical grade Sigma-Aldrich (USA), Merck (USA) and SRL (India) products of $>$ 99.9\% purity.

\textbf{Thermal imaging}
Thermal fluctuation of any surface (liquid, colloid or solid) immobilized to NC was determined using a FLIR-IR camera. A low power 658nm red laser source of 25mW was used to perturb the NC for the thermal measurements. The experiments were conducted in a temperature controlled enclosed room to avoid any unwanted thermal fluctuations which often hamper the measurements and start to calibrate the instrument.

\textbf{Flow cytometry}
Flow cytometry measurements were conducted in a BD-INFLUX flow system dedicated to nano particle science. The instrument is equipped with a multiple of five laser excitation sources and multiple detector channels to register the forward scattering (FSC), side scattering (SSC), polarization status and fluorescence emission of the samples under a flow cell. The optical polarization and depolarization status had been exploited in the context of SSC channels (38). Data analysis was performed using FlowJo V10 flow cytometry analysis software. Additional advantages of utilization of near infrared fluorescent (NIRF) window with optical filters (cut off, band pass or long pass filters) may improve the sensitivity and applicability of the instrument. Laser excitation channels of 488nm and 561nm wavelength had been utilized to illuminate the samples at a flow cell. Corresponding emissions were acquired at 692$\pm$40nm channel for 488nm laser and 670$\pm$30nm and 750nm LP (long pass channel) for 561nm excitation. 

\section{Results}
\textbf{AFM TEM and SEM of NC}
Topographical, ultra-structural and surface properties of the particles was measured using higher end atomic force microscopy (AFM), high-resolution transmission electron microscopy (HR-TEM) and scanning electron microscopy (SEM) respectively (Figure 1). 

\begin{figure}[ht]
	\includegraphics[width=10cm,angle=0]{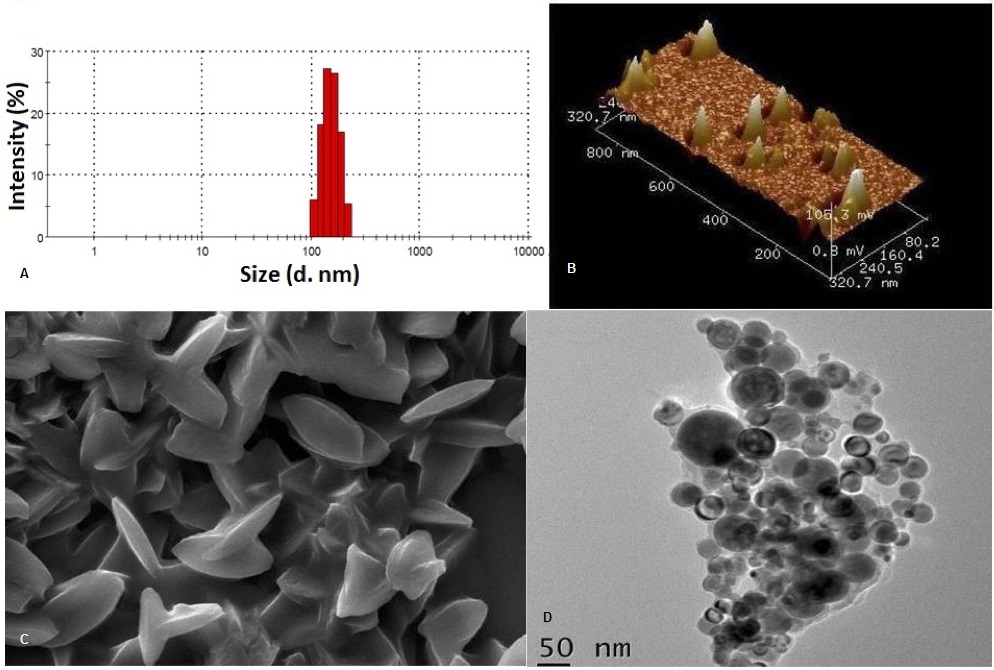}
	\caption{Nano scale characterization of NC: A, Hydrodynamic properties of NC dissolved in phosphate buffer (pH 7); B, 3-D height profile of the immobilized NC imaged by tapping mode atomic force microscope; C, Surface scanning image of the immobilized  andgold sputtered NC; D, Ultra-structural transmission electron microscopic analysis of NC.}
	\label{fig1}
\end{figure}

Semi-spherical and bean shaped particles bearing a size of 70nm with a three dimensional topography with uneven surface morphology had been confirmed by tapping mode atomic force measurements. 

Surface scanning microscopy (39) again correlate to the force microscopic results (Figure 1). Surface scanning experiments confirmed particulate nano-cone structure formed due to higher order self-assembly. The particles exhibit elongated bi-axial morphology with multi partite arrangement.

\begin{figure}[ht]
	\includegraphics[width=10cm,angle=0]{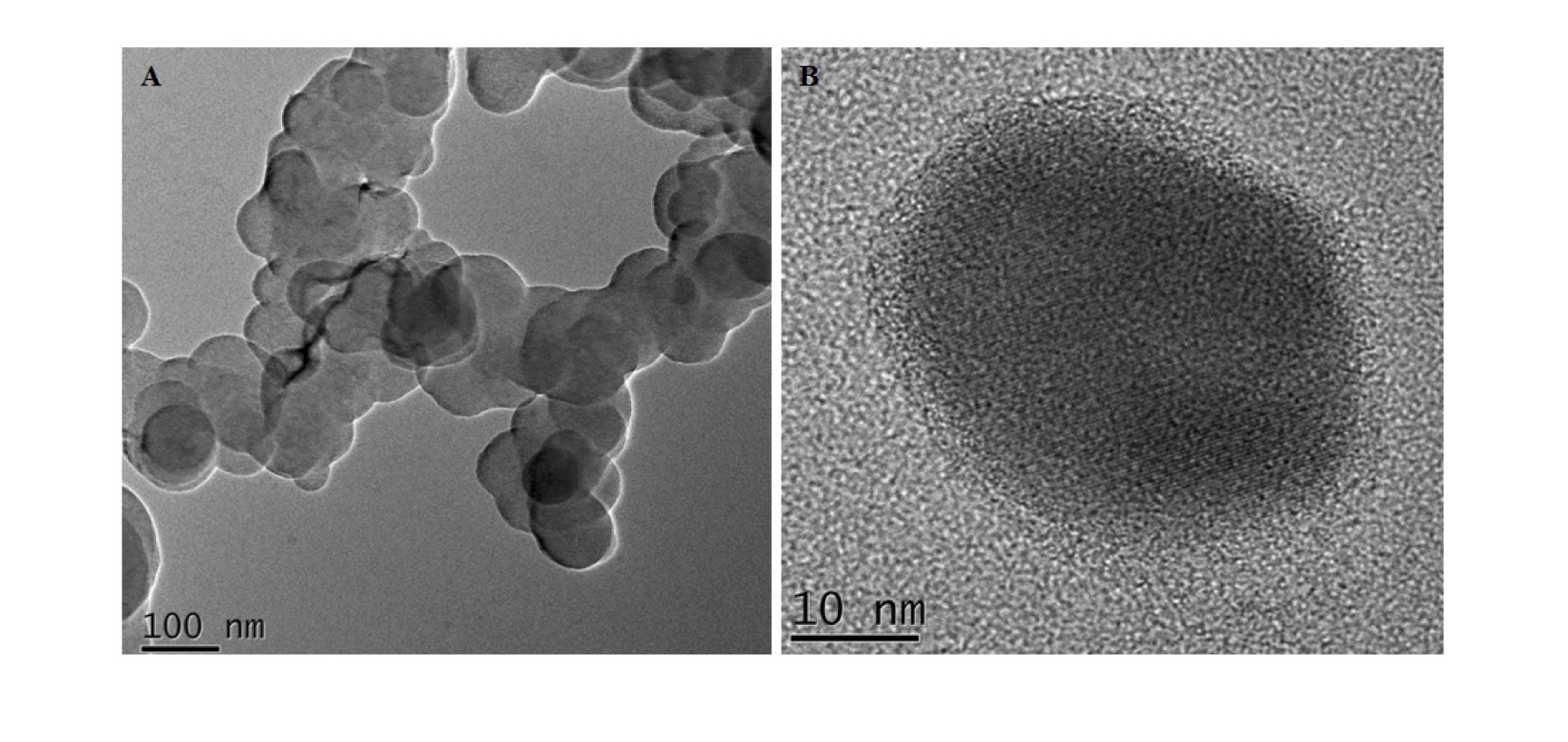}
	\caption{HR-TEM analysis of NC fixed at a copper grid. The scale bar is 100nm and 10nm for sub-figure A and B respectively.}
	\label{fig2}
\end{figure}

Higher resolution TEM images (HR-TEM) however decipher the actual arrangement of the self-assembled macro-cyclic backbones (Figure 2). Finely assembled self-stacked rings stabilized as higher order extensive domains consisting of ultra-structural patchy lattice structures are prominent. Ultra-structural and morphological fringe pattern may originate from self-assembled nano domains. TEM images of NC evidentially confirm the presence of particulate NC of 12-24 nm size.

textbf{Charge Distribution of NC} 
Self-assembly derived stabilizations of the photosynthetic dye chl $\textit{a}$ were found to correlate directly to the presence of appropriate stabilizing and capping conditions. A negative surface zeta potential of -20.6 mV signify the colloidal stability of the particles formed in water or any biological buffer such as phosphate buffer of pH 7. Precipitation of the macro-cycles due to un-controlled self-aggregation had been observed in water without any stabilizing agent. A critical concentration for the stabilizer citrate had been established. 12 to 15mM of citrate was found to be sufficient to stabilize NC with a hydrodynamic size of $~$86 to 100nm. 

\textbf{NIRF Fluorescence emissions from NC}
NIRF window had been tested for the colloidal assembly. A gain in the dual emission ratio ($P_{723}/P_{700}$) for NC implies a profound impact of nano scaling on the low energy emission band at 77K. Notably, a marked degree of stokes shifts had been observed for membrane bound chl $\textit{a}$ (Arachis hypogaea), free chl $\textit{a}$ (solvent extracted from Spenacea oleracea, Arachis hypogaea) and NC at cryogenic conditions. 

\textbf{SMF effects on 77K emission spectrum} 
Primarily, an external SMF source of 100 to 500mT strength was used to investigate the translation of magnetic transitions (41) to a photonic and spectroscopic output (fluorescence emission). Briefly the SMF effects were observed only at cryogenic temperatures (77K) and not at room temperature (Figure 3). In other words the SMF effect is absent in presence of thermal noise. The plausible inference that follows is that there is a strong dependence of SMF effect at particular assembly size that was stabilized. The difference of SMF effect on chl a and NC (panels A and D of Figure 3) further support this point. In the former case the SMF enhances the quantum yield in the NIRF peak, the reverse being true for NC. The lower decay rate of the NIRF at $\lambda_{max}$ = 723nm in presence of SMF (as compared to the same in absence of SMF) implies stabilization of the exited states and memory. 

\begin{figure}[ht]
	\includegraphics[width=10cm,angle=0]{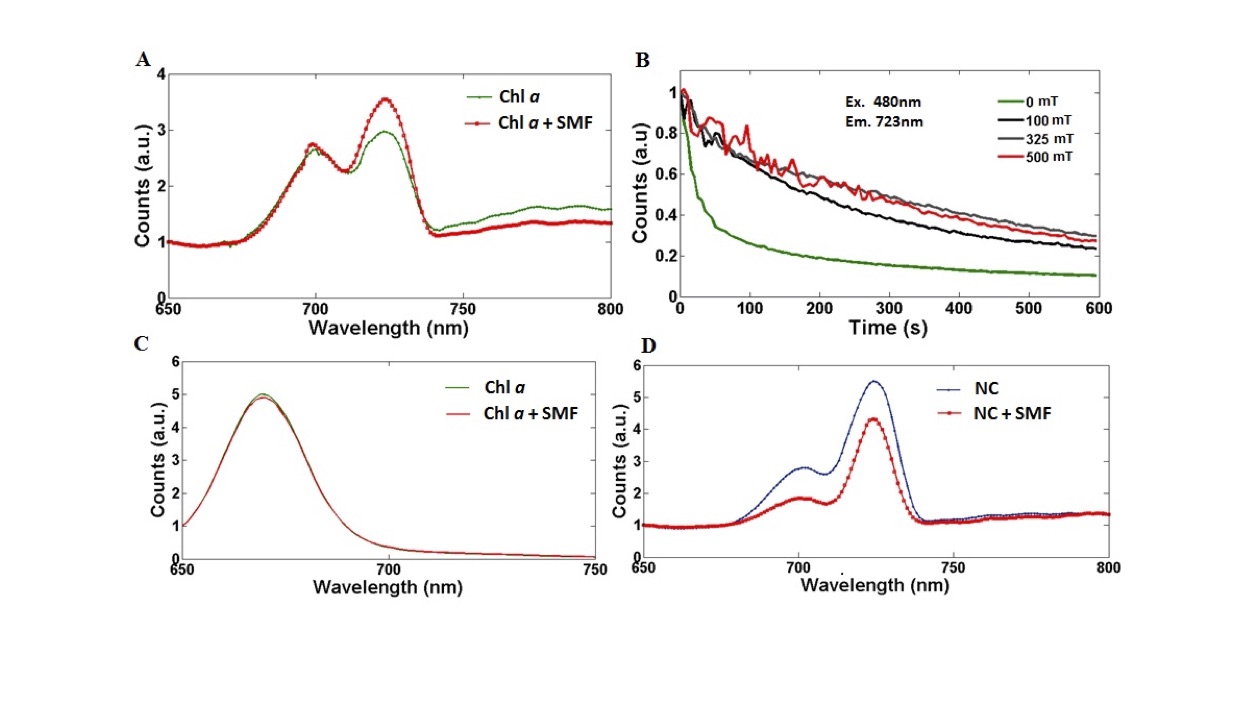}
	\caption{Fluoresence Spectroscopyexcitation at 430nm: A, 77 K emission spectrum -Dual peaks  of free chl $\textit{a}$ (green) and SMF exposed chl $\textit{a}$ (red) SMF enhancing the NIRF emission  B, Fluorescence decay kinetics of the NIRF peak  with increasing static magnetic field strengths showing lesser decay rates or higher stabilization of the excited states C, SMF effect at room temperatures; D, 77K spectra of NC in presence and absence of SMF - SMF attenuating the NIRF emission  (reciprocal to panel A - implying the SMF effect is assembly dependent) .}
	\label{fig3}
\end{figure}

\textbf{Fluorescence life time effects} 
Fluorescence lifetime decay analysis (Nano LED, Horiba scientific) of chl $\textit{a}$ and NC shows similar SMF response. Comparable lifetime values were reported previously for the free fluorophore at a solvent perturbed state (5.6 ns) as well as in buffer solutions (0.17-3ns) (42-44). Nano scale stabilizations of free chl $\textit{a}$ (3 ns) result in a substantial gain in the excited state lifetimes of NC to $>$ 4 ns values. Lifetime decay measurements (in the order of 10$^{-09}$ seconds) again affirm presence of magnetic memory (Table 1). 

\begin{table}[ht]
	\caption{Fluorescence lifetime decay analysis of chl $\textit{a}$ and NC at room temperatures. $\tau_{f}$ denote fluorescence decay time in the order of nanoseconds (ns).  $\tau_{f}$ values represent an average of at least two distinct experimental lifetime values. The lifetime values were derived from the multi-exponential fitting of the decay kinetic data points; Ex. and Em. denotes the excitation and emission maxima of the fluorophore and n/a denote not applicable.}
	\label{fl}
	\begin{tabular}{cclccl}
		\hline
		Fluorophore & Lifetime ($\tau_{f}$ , ns) & Ex. max (nm) & Em. max (nm) & Solvent\\
		\hline
		\hline
		Free Chla from spinach & 2.859	& 370 & 665 & acetone\\
		NC & 4.458 & 370	& 665	& phosphate buffer (pH 7.2)\\
		Free Chla & 4.37 & 370 & 665 & methanol \\
		Free Chla + SMF & 3.45 & 370 & 665 & acetone\\
		Free Chla + SMF & 3.56 & 370 & 665 & methanol\\
		NC + SMF & 3.18 & 370 & 665 & phosphate buffer (pH 7.2)\\
		\hline
		\hline
	\end{tabular}
\end{table}

Free chl $\textit{a}$ at acetone dissolved state exhibit a gain in fluorescence lifetime ($\tau_{f}$) upon SMF exposure. While free chl $\textit{a}$ dissolved in methanol exhibit a fast decay of $\tau_{f}$ after spin perturbations due to elevated secondary degradation reactions of chl $\textit{a}$ in methanol. However, $\tau_{f}$ was found to be less for NC (4.4.5ns) after SMF incubations (4.2.ns). A gain in the value of $\tau_{f}$ for acetone soluble free chl $\textit{a}$ and a fast decay of the excited state lifetime for NC upon SMF treatment may be attributed to as spin mixed excited state interactions (intra or inter), spin driven re-alignment or re-orientation or altered energy transfer dynamics or a nano scale spin quantum phenomena. Altered excited state fluorescence lifetimes for free chl $\textit{a}$ and NC again infer to a direct relation to the fluorescence decay rate kinetics, the size, shape, atomic packing, assembly pattern and stabilizations of the assemblies correlated to the structural, spectrophotometric and magnetic measurements. Notably, an external magnetic field sensing ability of such a nano assembly had been explored. Polarity screening based on hydrophobic self-interactions depending on the immediate surrounding solvent molecules may be utilized to design a green polarity meter.

\textbf{Magnetic properties of photosynthetic material} 
To address the problem of magnetic field sensing by such a green nano assembly (NC), and in search for a probable mechanistic basis for the SMF mediated spin interactions and memory, direct magnetic transitions of the samples was measured. Field dependent magnetization (M-H and M-T) curves for the nano particulate NC indicate profound magnetic transitions correlated to its nano scale ultra-structural properties, NIRF emissions at 77K and excited state lifetime analysis. M-H curves for NC at 300K, 150K and 5K indicate dominant diamagnetism for all of the curves with very typical nature (Figure 4). In the field region of $\pm$400 Oe, all the M-H curves show magnetic hysteresis with positive magnetization and magnetization increases with increasing field. In the field region above $\pm$400 Oe but below $\pm$2000 Oe, the M-H curves exhibit magnetic hysteresis with positive magnetization and magnetization decreases with increasing field. Further, in the field region above $\pm$2000 Oe but below $\pm$6000 Oe, the M-H curves show magnetic hysteresis with negative magnetization and magnetization decreases with increasing field. This kind of behavior of M-H curves above $\pm$400 Oe and below $\pm$6000 Oe is quite typical. This type of magnetic behavior manifested may be attributed to the presence of minuscule ferromagnetic domains, embedded in the highly diamagnetic matrix. Alternatively, a finite size dependency of the assembly at a nano range (superparamagnetism) may have adequate logic to address the problem.

\begin{figure}[ht]
	\includegraphics[width=10cm,angle=0]{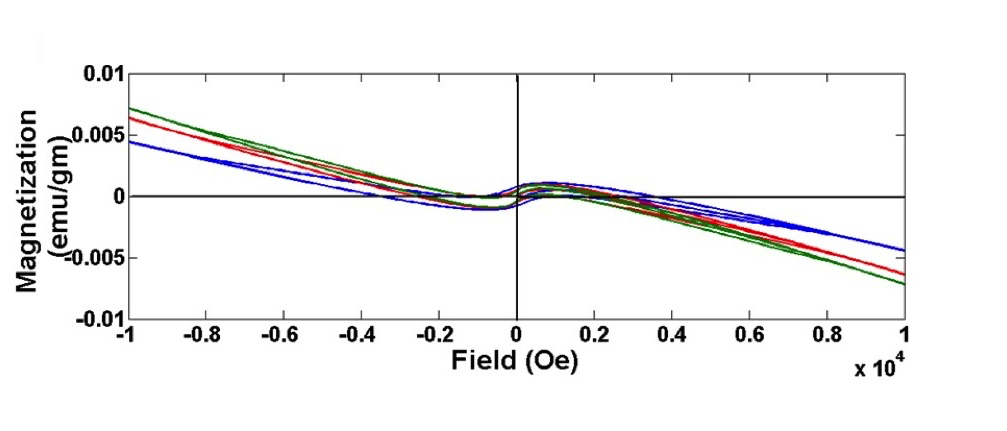}
	\caption{NC characterization: Magnetic hysteresis (M-H curve) Cyan, red and green lines represent magnetization at 5K, 150K and 300K respectively.}
	\label{fig4}
\end{figure}

Nano dimensional stabilization of the fluorophore results in a decrease of $M_{S}$ and an increase of H$_{C}$. Less M$_{S}$ at the nano scale may be accounted for either by a transition from multi to single domain or to a superparamagnetic range of particles or may be dependent on surface spin canting effects (45). For extreme small nano particles such kind of interactions may seem prominent but for larger particles these effects render insignificant. These results re-confirm the fractal nature of arrangement and ultra-structural stabilization of particles by higher order self-assembly. Uneven surface boundaries again may lead to generation of loose and disordered free electron spins at the particle surface other than particles net magnetic moment which often tend to cancellation of the net magnetization and lead to a lower value of M$_{S}$ in case of NC. However, the coercive force (H$_{C}$) was found to increase in case of NC with respect to the free state as with an extension of domain size from superparamagnetic (single-domain) to multi-domains, individual moment of each domain may not add up to orient in a particular direction. The M$_{S}$ can be improved to the bulk or free level by using different surface functionalization strategies to the particles which need to be explored further and is a subject for the future prospects of the work.
\begin{figure}[ht]
	\includegraphics[width=10cm,angle=0]{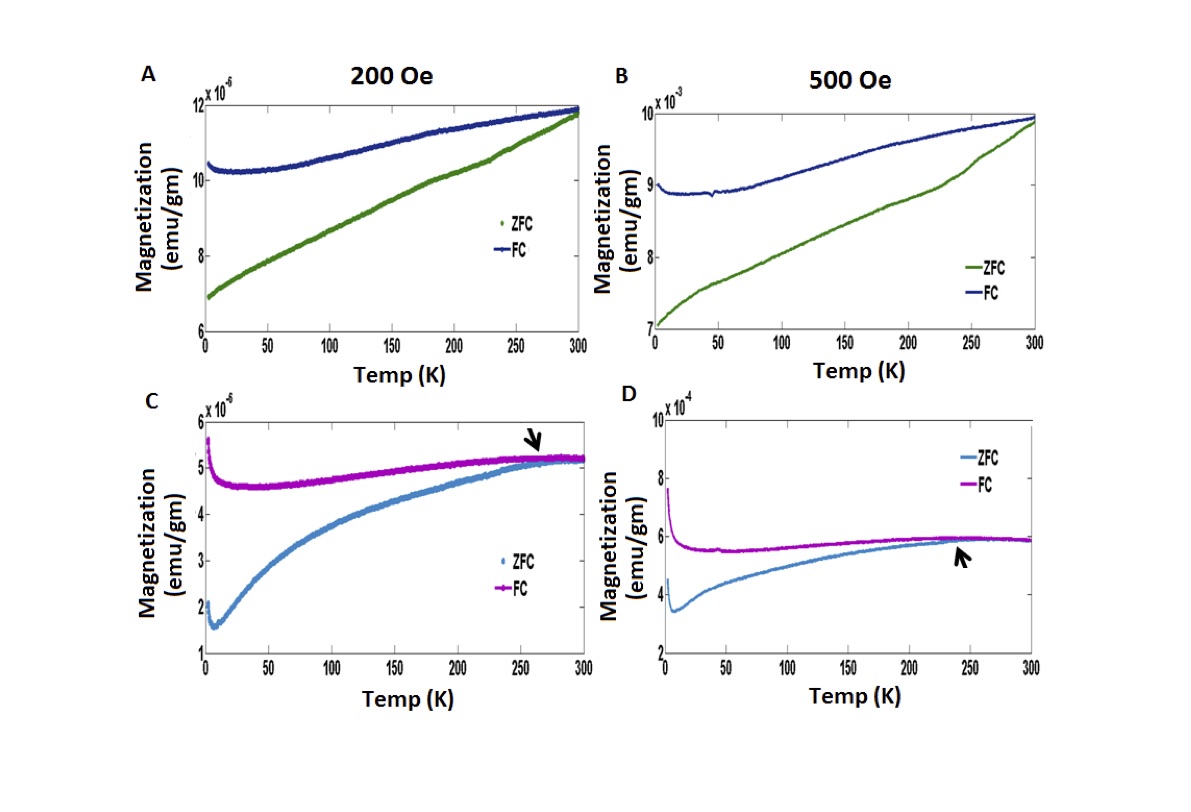}
	\caption{M-T measurements of the fluorophore and NC: Upper and lower panels illustrates the paired responses (zero field and with field) at two respective field strengths 200 Oe (left panel) and 500 Oe (right panel). Upper panel shows the response of chla and the lower panel illustrates the response of NC. Notably, the blocking temperature (T$_{B}$) was only attained for NC (arrow).}
	\label{fig5}
\end{figure}

Additionally, temperature dependent magnetization (M-T) measurements (Figure 5) under zero field cool (ZFC) and field cool (FC) conditions were performed for both the free and nano chl at 200 Oe and 500 Oe field strengths. The M-T curves under ZFC and FC conditions for NC initially exhibit a profound dip at very low temperature (below 10K) and then magnetization increases with increasing temperature. M-T curves under ZFC condition for the bulk or free chl however exhibit no such dip at low temperatures and in the entire temperature range magnetization increases with increasing magnetic field. Actually thermal energy acts as an agent of ordering energy of magnetism for the free and nano chl (except up to 10K). So far as the M-T curves under FC condition for the free chl is concerned both the curves at 200Oe and 500Oe indicate gradual increasing nature as a function of increase in temperature. Whereas for the nano assembled NC, both the curves at 200Oe and 500Oe indicated gradual decreasing nature with increase in temperature up to 100K and thereafter maintain a near constancy or rather a very small decreasing trend. The noteworthy feature in the M-T measurement is that the M-T curves under ZFC and FC conditions for the bulk chl do not intersect each other up to room temperature (300K). Whereas those same curves for the NC merges at around 250K. This type of nature of M-T curves is quite common for a superparamagnetic system. Stabilizations of NC may cause emergence of such nano scale magnetic property. In such a superparamagnetic system at blocking temperature (T$_{B}$), maximum value of magnetization was attained and M-T curves under ZFC and FC conditions intersect each other. Results are quite evident as $T_{B}$ for the bulk or free state is above 300K as M-T curves under ZFC and FC conditions do not collapse at 200Oe and even at 500Oe field strengths. Whereas in case of NC, maximum magnetization was attained and the M-T curves under ZFC and FC conditions converge at around 250K. Categorically, T$_{B}$ was attained only for the NC around 250K and no (or rather negligible) ferromagnetic interaction was present at 300K. Notably, NC exhibits profound and elevated magnetic transitions and memory limited by the size of the nano assembly with respect to the diamagnetic free fluorophore. 

\textbf{Some practical applications}
The nano-scale photosynthetic material can be subjected to a simple practical application. Using polarization enable flow cytometry we are able to classify different classes of carbon nano tube and graphene illustrated in Figure 6. The chlorophyll nanoparticle NC can rapidly discriminate and identify MWNT and graphene at a liquid and flow state by SSC-Pol and SSC-Depol phase analysis (complexity or granularity channel). Additionally, a liquid state sensing of nitric oxide had been examined. NC exert coherent heating by a low power (25mW) red laser of 658nm. Lastly, the response of NC to polarized light was measured which impart additional advantage to the nano NIRF tool.

\begin{figure}[ht]
	\includegraphics[width=8cm,angle=0]{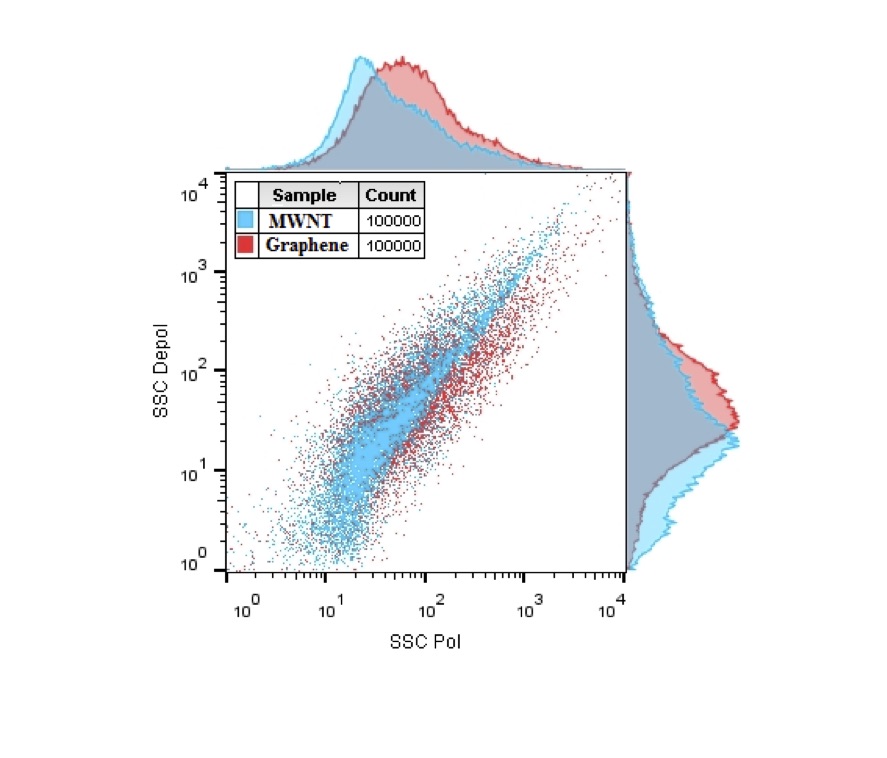}
	\caption{Applications of NC: Detection of MWNT (cyan) and graphene (red) by SSC-Pol (x-axis) and SSC-Depol (y-axis) at a flow cell at room temperatures.}
	\label{fig6}
\end{figure}
%

%

\section{Conclusions}
The chl a in non-polar solvents and NC in aqueous solvent show a varying range of spectral and magnetic properties with the former (spectral response) sensitive to presence of static magnetic field. The sensitivity to magnetic field suggests possibility of alignment of the porphyrin rings as the delocalized pi electrons therein can impart magnetic moment. Studies with apo-systems (porphyrin stripped) may provide some clue of the stated hypothesis and work in this direction is presently in progress. 

Lastly, Schr$\ddot{o}$dinger in his seminal book "What is life'' indicated that the negative entropy from the Sunlight as absorbed by the photoreceptor molecules may solve the thermodynamic riddle. As the higher entropic drive predicted by Boltzmann is apparently contradicted in the Darwin's theory which implies generation of order from disorder. The secret of the negative entropy seems to originate from the extreme high efficiency with which the light is converted into chemical energy. The 'quantum material' that does this job is chlorophyll. Magnetic characterization of this molecule naturally elevates the intricacy of the problem to new heights. It implies that the conversion may actually be integrated in nature. Interestingly, magnetic hysteresis of such a nano assembled bio probe has the advantage to address novel applications and methods development. Bio-compatible NC as a fluorescence (NIRF) enabled sensory tool may be exploited to address novel detection techniques, object characterizations, quantum measurements and as a flow cytometry probe. Quantum biological phenomena may be more common than what is known till today (e.g. quantum beats in photosynthesis or avian magneto-sensing). 
%

\section{acknowledgments}
The authors like to thank Department of Biotechnology, Govt. of India (BT/PR3957/NNT/28/659/2013) for providing funds. We thank DBT-CU-IPLS and CRNN-CU for providing high end infrastructural facilities and instruments and research. The authors show their gratitude to CoE, CU. We thank Dr. Maitree Bhattacharyya and Turban Kar for their support regarding the lifetime experiment. 

\section{abbreviations}
Chl $\textit{a}$, Chlorophyll $\textit{a}$; NC, nano chlorophyll;  MWNT, multi wall carbon nano tube; GSNO, S-nitroso-glutathione; DETA-NONOate, (Z)-1-(N-(2-aminoethyl)-N-(2-ammonioethyl) amino) diazen-1-ium-1,2-diolate; GNP, gold nanoparticle; SNP, silver nanoparticle; O.D, optical density; FACS, fluorescence-activated cell sorting; FSC, forward scattering; SSC, side scattering; FSC (par), parallel forward scattering; FSC (prep), perpendicular forward scattering; LP, long pass filter; mT, milli-tesla; Oe, oersted; $M_{S}$, saturation magnetization; $H_{C}$, coercivity; ZFC, zero field cool; FC, field cool; SMF, static magnetic field; NIRF, near infra-red fluorescence, $T_{B}$, blocking temperature.

\section{Reference}
1. A. W. Larkum, M. Kuhl: Chlorophyll d: the puzzle resolved. Trends Plant Sci 10 (8) 355-357.1. (2005).
doi:10.1016/j.tplants.2005.06.005

2. N. Murata: Control of excitation transfer in photosynthesis. ii. magnesium ion-dependent distribution of excitation energy between two pigment systems in spinach chloroplasts. Biochim. Biophys. Acta. Bioenergetics. 189 (2) 171-181 (1969). 
doi:10.1016/0005-2728(69)90045-0 

3. L. Fiedor, A. Kania, B. Mysliwa-Kurdziel, L. Orze l, G. Stochel: Understanding chlorophylls: central magnesium ion and phytyl as structural determinants. Biochim. Biophys. Acta. Bioenergetics. 1777 (12) 1491-1500 (2008). 
doi:10.1016/j.bbabio.2008.09.005

4. K. Ballschmiter, J. J. Katz: Infrared study of chlorophyll-chlorophyll and chlorophyll-water interactions. J. Am. Chem. Soc. 91 (10) 2661-2677 (1969). 
doi:10.1021/ja01038a044 

5. A. Agostiano, P. Cosma, M. Trotta, L. Monsu-Scolaro, N. Micali: Chlorophyll a behavior in aqueous solvents: formation of nanoscale self-assembled complexes. J. Phys. Chem. B. 106 (49) 12820-12829 (2002). 
doi:10.1021/jp026385k.

6. B. J. van Rossum, E. A. Schulten, J. Raap, H. Oschkinat, H. J. de Groot: A 3-d structural model of solid self-assembled chlorophyll a/h 2 o from multispin labeling and mas nmr 2-d dipolar correlation spectroscopy in  high magnetic field. J. Magn. Reson. 155 (1) 1-14 (2002). 
doi:10.1006/jmre.2002.2502. 

7. T. Miyatake, H. Tamiaki: Self-aggregates of natural chlorophylls and their synthetic analogues in aqueous media for making light-harvesting systems. Coord. Chem. Rev. 254 (21) 2593-2602 (2010). doi:10.1016/j.ccr.2009.12.027

8. D. Frackowiak, B. Zelent, H. Malak, A. Planner, R. Cegielski, R. Leblanc:  Fluorescence of aggregated forms of chl a in various media. J. Photochem. Photobiol., A  78 (1) 49-55 (1994). 
doi:10.1016/1010-6030(93)03707-n.

9. R. Vladkova: Chlorophyll a self-assembly in polar solvent-water mixtures. Photochem. Photobiol. 71 (1) 71-83 (2000). 
doi:10.1562/0031-8655(2000)0710071casaip2.0.co2

10. K. Sauer, J. R. L. Smith, A. J. Schultz: The dimerization of chlorophyll a, chlorophyll b, and bacteriochlorophyll in solution. J. Am. Chem. Soc. 88 (12) 2681-2688 (1966).
doi:10.1021/ja00964a011

11. N. Micali, A. Romeo, R. Lauceri, R. Purrello, F. Mallamace, L. M. Scolaro: Fractal structures in homo-and heteroaggregated water soluble porphyrins. J. Phys. Chem. B. 104 (40) 9416-9420 (2000).
doi:10.1021/jp001750y

12. C. J. Medforth, Z. Wang, K. E. Martin, Y. Song, J. L. Jacobsen, J. A. Shelnutt: Self-assembled porphyrin nanostructures. Chem. Commun.  (47) 7261-7277 (2009).
doi:10.1039/b914432c

13. A. P. Gerola, T. M. Tsubone, A. Santana, H. P. de Oliveira, N. Hioka, W. Caetano: Properties of chlorophyll and derivatives in homogeneous and  microheterogeneous systems. J. Phys. Chem. B. 115 (22) 7364-7373 (2011).
doi:10.1021/jp201278b

14. A. D. Schwab, D. E. Smith, B. Bond-Watts, D. E. Johnston, J. Hone, A. T. Johnson, J. C. de Paula, W. F. Smith: Photoconductivity of self-assembled porphyrin nanorods. Nano Lett.  4 (7) 1261-1265 (2004)
doi:10.1021/nl049421v

15. A. L. Yeats, A. D. Schwab, B. Massare, D. E. Johnston, A. T. Johnson, J. C. de Paula, W. F. Smith: Photoconductivity of self-assembled nanotapes made from meso-tri (4-sulfonatophenyl) monophenylporphine. J. Phys. Chem. 112 (6) 2170-2176 (2008).
doi:10.1021/jp0765695

16. M. Jurow, A. E. Schuckman, J. D. Batteas, C. M. Drain: Porphyrins as molecular electronic components of functional devices. Coord. Chem. Rev. 254 (19) 2297-2310 (2010).
doi:10.1016/j.ccr.2010.05.014

17. A. D. Schwab, D. E. Smith, C. S. Rich, E. R. Young, W. F. Smith, J. C. de Paula: Porphyrin nanorods. J. Phys. Chem. B 107 (41) 11339-11345 (2003).
doi:10.1021/jp035569b

18. W. Tu, Y. Dong, J. Lei, H. Ju: Low-potential photoelectrochemical biosensing using porphyrin-functionalized tio2 nanoparticles. Anal. Chem. 82 (20) 8711-8716 (2010).
doi:10.1021/ac102070f

19. M. R. Wasielewski: Self-assembly strategies for integrating light harvesting and charge separation in artificial photosynthetic systems. Acc. Chem. Res. 42 (12) 1910-1921 (2009).
doi:10.1021/ar9001735

20. D. Monti, S. Nardis, M. Stefanelli, R. Paolesse, C. Di Natale, A. D'Amico: Porphyrin-based nanostructures for sensing applications. Journal of Sensors (2009).
doi:10.1155/2009/856053

21. M. Alcoutlabi, G. B. McKenna: Effects of confinement on material behaviour at the nanometre size scale. J. Phys. Condens. Matter. 17 (15) R461 (2005).
doi:10.1088/0953-8984/17/15/r01

22. K. Seo, K. Varadwaj, P. Mohanty,  S. Lee, Y. Jo, M.-H. Jung, J. Kim, B. Kim: Magnetic properties of single-crystalline cosi nanowires. Nano lett. 7 (5) 1240-1245 (2007).
doi:10.1021/nl070113h

23. S.-W. Hung, T. T.-J.Wang, L.-W. Chu, L.-J. Chen: Orientation-dependent room-temperature ferromagnetism of fesi nanowires and applications in nonvolatile memory devices. J. Phys. Chem. 115 (31) 15592-15597 (2011).
doi:10.1021/jp201395r

24. Y.-H. Chi, L. Yu, J.-M. Shi, Y.-Q. Zhang, T.-Q. Hu, G.-Q. Zhang, W. Shi, P. Cheng: π-π stacking and ferromagnetic coupling mechanism on a binuclear cu (ii) complex. Dalton Transactions 40 (7) 1453-1462 (2011).
doi:10.1039/c0dt01127d

25. M. Ding, B. Wang, Z. Wang, J. Zhang, O. Fuhr, D. Fenske, S. Gao: Constructing single-chain magnets by supramolecular stacking and spin canting: A case study on manganese (iii) corroles. Chem. Eur. J. 18 (3) 915-924 (2012).
doi:10.1002/chem.201101912

26. C. E. Schulz, H. Song, Y. J. Lee, J. U. Mondal, K. Mohanrao, C. A. Reed, F. A. Walker, W. R. Scheidt: Metalloporphyrin. pi.-cation radicals. molecular structure and spin coupling in a vanadyl octaethylporphyrinate derivative, an unexpected spin coupling path. J. Am. Chem. Soc.  116 (16) 7196-7203 (1994).
doi:10.1021/ja00095a023

27. K. Kano, H. Minamizono, T. Kitae, S. Negi: Self-aggregation of cationic porphyrins in water. can pi pi stacking interaction overcome electrostatic repulsive force?. J. Phys Chem. A 101 (34) 6118-6124 (1997).
doi:10.1021/jp9710446

28. Z. Wang, C. J. Medforth, J. A. Shelnutt: Porphyrin nanotubes by ionic self-assembly. J. Am. Chem. Soc. 126 (49) 15954-15955 (2004).
doi:10.1021/ja045068j

29. J. C. de Paula, J. H. Robblee, R. F. Pasternack: Aggregation of chlorophyll a probed by resonance light scattering spectroscopy. Biophys. J.  68 (1) 335 (1995).
doi:10.1016/s0006-3495(95)80192-x

30. J. Wang, H. Duan, Z. Huang, B. Karihaloo: A scaling law for properties of nano-structured materials. Proc. R.  Soc. A 462 (2069) 1355-1363 (2006).
doi:10.1098/rspa.2005.1637

31. Y.-w. Jun, J.-w. Seo, J. Cheon: Nanoscaling laws of magnetic nanoparticles and their applicabilities in biomedical sciences. Acc. Chem. Res.  41 (2) 179-189 (2008).
doi:10.1021/ar700121f

32. H. Murai: Spin-chemical approach to photochemistry: reaction control by spin quantum operation. J. Photochem. Photobiol. 3 (3) 183-201 (2003).
doi:10.1016/s1389-5567(02)00038-2

33. H. T. Quach, R. L. Steeper, G. W. Griffin: An improved method for the extraction and thin-layer chromatography of chlorophyll a and b from spinach. J. Chem. Educ. 81 (3) 385 (2004).
doi:10.1021/ed081p385

34. Abhishek Bhattacharya, Anjan Kr. Dasgupta: Chlorphyll nnanoparticle and process for its production. Patent IPR/FA/13019 (2014).

35. G. Frens: Controlled nucleation for the regulation of the particle size in  monodisperse gold suspensions. Nat. 241 (105) 20-22 (1973).
doi:10.1038/physci241020a0

36. S. Singha, H. Datta, A. K. Dasgupta: Size dependent chaperon properties of gold nanoparticles. J. Nanosci. Nanotechnol. 10 (2) 826-832 (2010).
doi:10.1166/jnn.2010.1805

37. T. Bhattacharyaa, A. K. Dasgupta: A thermodynamic discriminator for carbon nanomaterials. arXiv 1507.0.1672 (2015).

38. A. K. Dasgupta, T. Bhattacharyya, S. Roy: Configurational chirality based separation. US Patent App. 14/539,790 (2014).
doi:10.1039/c4ta00110a

39. S. Thiberge, A. Nechushtan, D. Sprinzak, O. Gileadi, V. Behar, O. Zik, Y. Chowers, S. Michaeli, J. Schlessinger, E. Moses: Scanning electron microscopy of cells and tissues under fully hydrated conditions. Proc. Natl. Acad. Sci.  101 (10) 3346-3351 (2004).
doi:10.1073/pnas.0400088101

40. A. Bhattacharya, M. Chakraborty, S. O. Raja, A. Ghosh, M. Dasgupta, A. K. Dasgupta: Static magnetic field (smf) sensing of the p 723/p 689 photosynthetic complex. Photochem. Photobiol. Sci. 13,1719-1729 (2014).
doi:10.1039/c4pp00295d

41. J. R. Norris: Triplet states and photosynthesis. Photochem. Photobiol. 23 (6) 449-450 (1976).
doi:10.1111/j.1751-1097.1976.tb07278.x

42. M. Kaplanova, L. Parma: Effect of excitation and emission wavelength on the fluorescence lifetimes of chlorophyll a. Gen Physiol Biophys. 3, 127-134 (1984).
doi:10.1016/0047-2670(81)80005-6

43. M. Y. Berezin, S. Achilefu: Fluorescence lifetime measurements and biological imaging. Chem. Rev. 110, 2641-2684 (2010).
doi:10.1021/cr900343z

44. T. J. Bensky, L. Clemo, C. Gilbert, B. Neff, M. A. Moline, D. Rohan: Observation of nanosecond laser induced fluorescence of in vitro seawater phytoplankton. Appl. Opt.  47, 3980-3986 (2008).
doi:10.1364/ao.47.003980

45. M. Morales, S. Veintemillas-Verdaguer, M. Montero, C. Serna, A. Roig, L. Casas, B. Martinez, F. Sandiumenge: Surface and internal spin canting in gama-fe2o3 nanoparticles. Chem. Mater. 11 (11) 3058-3064 (1999).
doi:10.1021/cm991018f

%
%
%

\end{document}